\documentclass[letterpaper, 10 pt, conference]{ieeeconf}  

\IEEEoverridecommandlockouts                              

\usepackage{graphics} 
\usepackage{graphicx}
\usepackage{caption}
\usepackage{subcaption}
\usepackage{epsfig} 
\usepackage{mathptmx} 
\usepackage{times} 
\usepackage{amsmath} 
\usepackage{amssymb}  

\usepackage{makecell}
\usepackage{float}
\usepackage{tikz}
\usetikzlibrary{positioning}
\usetikzlibrary{shapes.geometric, arrows}

\title{\LARGE \bf
Automated Vehicle Highway Merging: Motion Planning via Adaptive Interactive Mixed-Integer MPC
}

\author{Viranjan Bhattacharyya$^{1}$ and Ardalan Vahidi
\thanks{\scriptsize *This work was supported by Argonne SMART 2.0}
\thanks{ \scriptsize Viranjan Bhattacharyya ({\tt vbhatta@clemson.edu}) and Ardalan Vahidi ({\tt avahidi@clemson.edu}) are with the Department of Mechanical Engineering, Clemson University, Clemson, SC, USA}
\thanks{\scriptsize $^{1}$Corresponding Author}
}

\begin{document}

\maketitle
\thispagestyle{empty}
\pagestyle{empty}

\begin{abstract}
A new motion planning framework for automated highway merging is presented in this paper. To plan the merge and predict the motion of the neighboring vehicle, the ego automated vehicle solves a joint optimization of both vehicle costs over a receding horizon. The non-convex nature of feasible regions and lane discipline is handled by introducing integer decision variables resulting in a mixed integer quadratic programming (MIQP) formulation of the model predictive control (MPC) problem. Furthermore, the ego uses an inverse optimal control approach to impute the weights of neighboring vehicle cost by observing the neighbor’s recent motion and adapts its solution accordingly. We call this adaptive interactive mixed integer MPC (aiMPC). Simulation results show the effectiveness of the proposed framework.

\end{abstract}

\section{INTRODUCTION}
One of the major challenges in autonomous driving is anticipation of neighboring vehicles’ (NV) behavior. The actions of the ego vehicle affect and are affected by the actions of neighboring vehicles \cite{fisac2019hierarchical}. This mutual influence of driving behavior is called \textit{interaction}.

The motion planning problem in vehicle merging is particularly a problem where the coupling is significant as there is a strong mutual influence between the ego vehicle and its neighbor(s). Unilateral `pipeline' based approaches do not consider the ability of the ego to influence the trajectory of the NV. Common unilateral approaches \cite{sciarretta2020perception} include assumption of constant speed, constant acceleration, speed-dependent acceleration or intention sharing by NV through V2V connectivity \cite{dollar2018predictively}. The deviations from reality may be handled through probabilistic models \cite{havlak2013discrete, dollar2019automated}. However, these approaches tend to be conservative since the motion plans are computed based on models of NV as if it acts in isolation. On the other hand, intention sharing through V2V communication involves practical challenges. The NV may not follow its shared intentions and it may be impractical in highway scenarios where intention sharing may not be possible due to the lack of network availability.
Authors in some of the previous research work have recognized this limitation of traditional motion planning approaches. \cite{trautman2010unfreezing} presented a joint collision avoidance approach for robots. \cite{litkouhi_wei_dolan_2014} presents an interactive merge management system with a relatively coarse decision space. It considers only the longitudinal control of the ego vehicle and does not incorporate system constraints explicitly.

Inspired by recent research \cite{schwarting2019social, sadigh2018planning}, in this paper we present an optimal control and inverse optimal control based method for \textit{interactive} motion planning in automated driving highway merging. Our method not only incorporates the interactions, but also adapts the joint optimization cost function based on observation of the actual trajectory of the NV, online. Our hypothesis is that the nature of the NV is defined by weight it places on various terms in its cost function. These weights need to be estimated online through trajectory observations. To facilitate this, we harness the inverse optimization theory presented in \cite{keshavarz2011imputing}.

\begin{figure}[h!]
    \centering
    \includegraphics[width=0.5\textwidth]{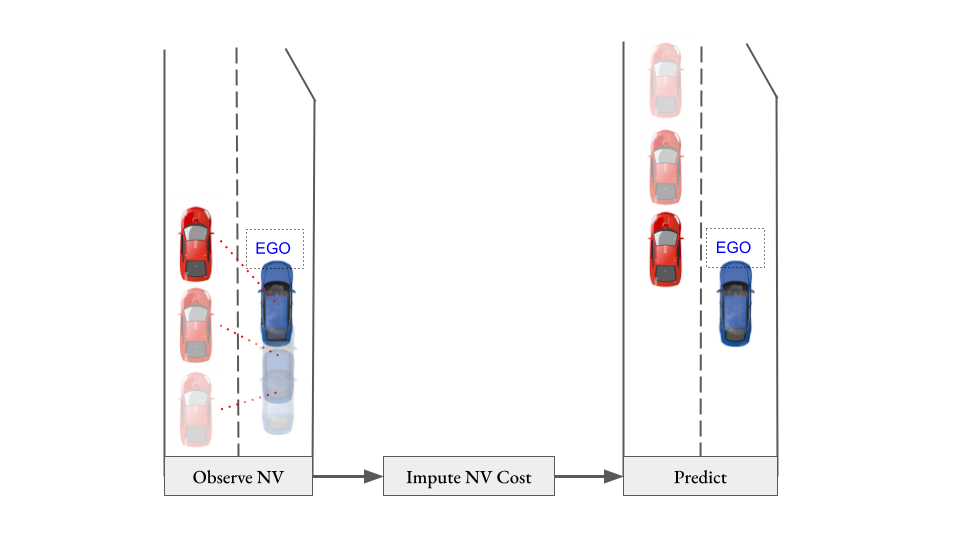}
    \caption{Schematic diagram of the proposed framework}
    \label{schematic}
\end{figure}

Figure \ref{schematic} visually represents our method when there is a single NV. The ego (blue) vehicle observes the immediate NV (red) for a few time instants and using the observed trajectory, imputes a cost function. It then utilizes this imputed cost function in a joint mixed integer MPC resulting in an interactive predictive control strategy. Our proposed framework requires only the current states of the NV which may be obtained either via V2V connectivity or on-board sensors. This makes the approach applicable to scenarios involving unconnected vehicles and out of network-coverage areas.

The paper is organized as follows: Section \ref{interaction} presents the interactive joint MPC problem formulation. The simulation model of the NV is also presented.
Next, in section \ref{estimation} we briefly discuss the inverse optimization theory for online estimation of NV's cost function and its implementation in the paper. Finally, in section \ref{simulation} we present simulation results that show the effectiveness of our proposed framework in online estimation of NV's behavior, interactive motion planning and executing safe merging.

\section{INTERACTION MODELING}\label{interaction}

 In order to capture the \textit{interaction}, we formulate the motion planning problem as a joint mixed-integer quadratic program (MIQP), solved in a receding horizon fashion via MPC. The cost terms associated with the NV are adapted online and therefore, we call this adaptive interactive mixed-integer MPC (aiMPC).
 
 \begin{subequations}\label{aiMPC formulation}
    \begin{gather}
    \begin{split}
    \min_{\tiny{u_{ego}, u_{NV}, \sigma_{logic}}}\sum_{i=k}^{k+N-1} J_{ego}(x_{ego}(i),u_{ego}(i))
    \\ +\; J_{NV}(\alpha,x_{NV}(i),u_{NV}(i)) + J^{Terminal}
    \end{split}
    \\s.t.\;\;\;\;\;x_{ego}(i+1)=A_{d} x_{ego}(i) + B_{d} u_{ego}(i)
    \\\;\;\;\;\;x_{NV}(i+1)=A_{d,NV} x_{NV}(i) + B_{d,NV} u_{NV}(i)
    \\x_{vehicle}(i)\in \mathrm{X}_{free}
    \\u_{vehicle}(i) \in \mathrm{U}
    \\\sigma_{logic}(i) \in \{0,1\}
    \end{gather}
\end{subequations}

where, $J_{vehicle}, \; x_{vehicle}, \; u_{vehicle}$ and $\sigma_{logic}$ are the quadratic cost, states, controls and logical variables respectively with $vehicle \in \{ego, NV\}$. $\alpha$ is the vector of weights estimated online, $X_{free}$ is the safe region to avoid collision and $U$ is the set of admissible controls.

The aiMPC formulation is based on \cite{dollar2018predictively}. However, unlike \cite{dollar2018predictively}, in this paper we do not assume that the intention of the NV is shared with the ego. Instead, the joint cost, dynamics and collision avoidance constraints predict the trajectory of the ego and NV - capturing the mutual influence over the MPC horizon. The states of the NV are re-initialized according to the actual observation at every time step. This acts as feedback to the controller.

\subsection{The Cost Function}
The cost function of the aiMPC includes the cost function of the ego as well as the NV. In this work, we limit our analysis to one NV, but the framework can be extended to multiple immediate NVs. The cost function used here assumes that the vehicles have already been assigned a driving schedule which translates to reference velocity and position. This is consistent with the literature \cite{rios2016survey} and involves minimizing the difference between the actual and reference state.

\begin{subequations}
\begin{gather}
    \begin{split}
    J_{ego} = q_{s,ego}(s_{ego} - s_{ref})^2 + q_{v,ego}(v_{ego} - v_{ref})^2 + \\ q_{a,ego}a_{ego}^2 + q_{l,ego}(l_{ego} - l_{ref})^2 + q_{a,ego}u_{a,ego}^2
    \end{split}
    \\ J_{NV} = \alpha_s(s_{NV} - s_{ref,NV})^2 + \alpha_v(v_{NV} - v_{ref,NV})^2 + \alpha_a a_{NV}^2
\end{gather}
\end{subequations}

\subsection{Vehicle Model}
Since this is a high level motion planner, consistent with the state of the art, we use a moderate fidelity kinematic model. 

\subsubsection{Ego}
The model we use for the ego vehicle was developed by Dollar et al \cite{dollar2018predictively} and decouples the lateral and longitudinal dynamics.

\begin{equation}
\small{
\frac{d}{dt}
\begin{bmatrix}
    s\\v\\a\\l\\r_l
\end{bmatrix}
=
\begin{bmatrix}
    0 & 1 & 0 & 0 & 0\\
    0 & 0 & 1 & 0 & 0\\
    0 & 0 & -\frac{1}{\tau} & 0 & 0\\
    0 & 0 & 0 & 0 & 1\\
    0 & 0 & 0 & -\omega_n^2 & -2\zeta\omega_n
\end{bmatrix}
\begin{bmatrix}
    s\\v\\a\\l\\r_l
\end{bmatrix}
+
\begin{bmatrix}
    0 & 0\\
    0 & 0\\
    \frac{1}{\tau} & 0\\
    0 & 0\\
    0 & K\omega_n^2
\end{bmatrix}
\begin{bmatrix}
    u_a\\
    u_l
\end{bmatrix}
}
\end{equation}

The vehicle's longitudinal motion is considered as a double integrator with the states being position, velocity and acceleration, $[s,v,a]$, and time constant $\tau$, while the lateral motion is cast as a second-order critically damped system with $[l,r_l]$ being the lateral lane position and rate of change of lane respectively. $\zeta$ is the damping ratio, $K$ is the gain and $\omega_n$ is the natural frequency. The control inputs are longitudinal acceleration $u_a$ and lane command $u_l$. It may be specified that $u_l$ is an integer equal to the commanded lane number.
We discretize these dynamics when implementing in the controller as $x_{ego}(k+1) = A_d x_{ego}(k) + B_d u_{ego}(k)$ where $x_{ego} = [s\;v\;a\;l\;r_l]^T$ and $u_{ego} = [u_a\;u_l]^T$.

\subsubsection{NV}
The dynamics of the NV (used in joint optimization by ego) are modeled as,

\begin{equation}
\frac{d}{dt}
    \begin{bmatrix}
        s_{NV}\\v_{NV}\\a_{NV}
    \end{bmatrix}
    =
    \begin{bmatrix}
        0 & 1 & 0\\
        0 & 0 & 1\\
        0 & 0 & -\frac{1}{\tau}
    \end{bmatrix}
    \begin{bmatrix}
        s_{NV}\\v_{NV}\\a_{NV}
    \end{bmatrix}
    +
    \begin{bmatrix}
        0\\0\\\frac{1}{\tau}
    \end{bmatrix}
    u_{NV}
\end{equation}

These continuous time dynamics are discretized as $x_{NV}(k+1) = A_{d,NV} x_{NV}(k) + B_{d,NV} u_{NV}(k)$ where $x_{NV} = [s\;v\;a]^T$ and $u_{NV}$ is the longitudinal acceleration control. In this work we assume that the NV drives in its own lane.

\begin{table}[h!]
\centering
\caption{Model parameters}
\begin{tabular}{c c} 
 \hline
 Parameter &  Value \\ [0.5ex] 
 \hline\hline
 $\tau$ & 0.275 s \\[0.5ex]
 \hline
 $\omega_n$ & 1.091 rad/s\\[0.5ex]
 \hline
 $\zeta$ & 1\\[0.5ex]
 \hline
 $K$ & 1\\
 \hline
\end{tabular}
\label{model params}
\end{table}

 \subsection{Joint Collision Avoidance Constraints}
The drive-able region $X_{free}$ is non-convex. If in the same lane as NV, the ego vehicle either needs to be ahead of the front of NV or behind the rear of NV. We utilize the Big M method \cite{vecchietti2003modeling} to convert this \textit{OR} constraint to \textit{AND}, making it suitable for mixed integer programming.


\begin{subequations}\label{collision avoid}
    \begin{gather}
        s_{ego} - s_{NV} - M\beta_{l}^{NV} - M\mu_{l} \geq (L + gap) - 2M \label{ahead}\\
        s_{NV} - s_{ego} + M\beta_{l}^{NV} - M\mu_{l} \geq (L + gap) - M \label{behind}
    \end{gather}
\end{subequations}


In equation (\ref{collision avoid}), we introduce the logical decision variables $\beta_{l}^{NV}$ and $\mu_{l}$. $\beta_{l}^{NV}$ is the front-rear indicator. The choice of $1$ places ego ahead of the NV in lane $l$ and $0$ places it behind.  $\mu_{l}$ is the lane indicator. If set to $1$, ego resides in the lane $l$ and if set to $0$ it does not. In effect, the equation (\ref{ahead}) ensures that ego maintains a distance equal to the vehicle length $L$ plus a desired $gap$ ahead of the NV when it is placed in the same lane and ahead of NV. When this is the case, (\ref{behind}) gets inactive. On the other hand, when ego resides behind NV, (\ref{behind}) ensures that it maintains $L+gap$ behind it while (\ref{ahead}) gets inactive.

\subsection{Control Admissibility}
The acceleration control admissibility constraints combine velocity with the acceleration command to prevent operation in mechanically infeasible regions \cite{dollar2017quantifying}.

\begin{subequations}\label{control admissibility}
    \begin{gather}
        u_{a} \geq u_{a,min}
        \\ u_{a} \leq m_1 v + b_1
        \\ u_{a} \leq m_2 v + b_2
    \end{gather}
\end{subequations}
Here, the maximal acceleration is velocity dependent and (\ref{control admissibility}) is a convex approximation to the feasible region.

\subsection{Simulating the Neighbor}
For this work, we simulate the NV as controlled by an MPC which solves the following quadratically constrained quadratic program (QCQP)

\begin{subequations}
    \begin{gather}
    \min_{u_{NV}^{actual}}\sum_{i=k}^{k+T-1} J_{NV}(q, x_{NV}(i),x_{ego}(i),u_{NV}^{actual}(i))
    \\s.t.\;\;\;\;\;x_{NV}(k+1) = A_{d,NV} x_{NV}(k) + B_{d,NV} u_{NV}^{actual}(k)
    \\\;\;\;\;\;x_{ego}(i+1)=g_{projection}(x_{ego}(i))
    \\x_{NV}(i)\in \mathcal{X}_{free}
    \\u_{NV}^{actual}(i) \in \mathcal{U}
    \\ \mathcal{X}_{free} \doteq \dfrac{(s_{NV}(i) - s_{ego}(i))^2}{a^2} + \dfrac{(l_{NV}(i) - l_{ego}(i))^2}{b^2} \geq 1
    \end{gather}
\end{subequations}

Physically, this translates to the NV projecting the ego vehicle into the future to predict its trajectory over the horizon via the $g_{projection}()$ function. This function determines the future longitudinal and lateral position of the ego from the current observed states. In this work, we assume that the NV projects the ego by holding the current longitudinal ($v$) and lateral ($r_l$) velocity constant over the horizon. The NV controller also maintains a safety distance from the ego by ensuring that it lies outside an ellipse encapsulating the ego with semi-major axis $a$, semi-minor axis $b$ and centered on the ego. Other expressions have the same meaning as before while it may be noted that $u_{NV}^{actual}$ is the actual control applied by NV which is not available to the ego.

\section{ESTIMATING NEIGHBOR BEHAVIOR}\label{estimation}
Inspired by previous research \cite{schwarting2019social}, we aim to estimate the behavior of the neighboring vehicle (NV) online by observing its trajectory. In the present problem, we define \textit{behavior} based on weights associated with various terms in the cost function of the NV. 
Our approach imputes \cite{keshavarz2011imputing} these cost parameters (weights) based on the observed trajectory (data).

\subsection{Background}
In order to keep the paper reasonably self contained, we give a brief background on imputing a convex objective function from data. The detailed theory can be found in \cite{keshavarz2011imputing}.

The aim is to fit a convex forward optimization problem (\ref{forward problem}) to the observed trajectory data,
\begin{subequations}\label{forward problem}
   \begin{gather}
       \min_{x} f(x)
       \\ s.t. \; h(x,u) = 0
       \\ g_j(x,u) \leq 0 \; for j = 1,...,p
   \end{gather}
\end{subequations}
where $x \in \mathbb{R}^n$ are the state data, $u \in \mathbb{R}^m$ are the controls, $f(x)$ is a convex function, $h(x,u)$ is linear in $x$ and $u$ and $g_j$ are convex.
Assuming that the observed trajectory is \textit{approximately optimal} with respect to problem (\ref{forward problem}), we define residuals based on the Karush–Kuhn–Tucker (KKT) conditions. These residuals are deviations from the first order necessary conditions of optimality.
\begin{subequations}
   \begin{gather}
   r_{eq} = h(x,u)
   \\ r_{ineq} = (g_j(x,u))_{+}
   \\ r_{stat}(\alpha, \lambda, \nu) =  \nabla f(x) + \Sigma_{j=1}^{p} \lambda_j \nabla g_j(x,u) + \nu h(x,u)
   \\ r_{comp}(\lambda) = \lambda_j g_j(x,u), \;\;\; j=1,...,p
   \end{gather}
\end{subequations}
Here, $r_{eq}$ and $r_{ineq}$ are the primal feasibility conditions which are independent of the dual variables. We assume they are close to zero. Now, for the observed trajectory to be \textit{approximately optimal}, the residuals $r_{stat}$ and $r_{comp}$ are minimized. The dual variables $\lambda, \nu$ and the imputation parameter $\alpha$ are the decision variables and
\begin{subequations}
   \begin{gather}
   f = \sum_{j} \alpha_j f_j, \;\;\; \alpha = [\alpha_1,...,\alpha_j]
   \end{gather}
\end{subequations}
where $f_j$ are the basis functions of the cost function. Our goal is to find $\alpha$ for which $f$ is (approximately) consistent with the observed trajectory data.

\subsection{Formulation}
We fit the following optimization problem to the observed trajectory.
\begin{subequations}\label{fit problem}
    \begin{gather}
        \min_{x_{NV}^{(i+1)}} \sum_{i=k-r}^{k-1}  J_{NV}(\alpha, x_{NV}^{(i+1)})
        \\s.t. \;\;\; h(x_{NV}^{(i+1)},x_{NV}^{(i)},u_{NV}^{(i)}) = 0
        \\ g_j(x_{NV}^{(i+1)},x_{Ego}^{(i+1)},u_{NV}^{(i)}) \leq 0, \;\;\; j = 1,...,p
    \end{gather}
\end{subequations}
where, $k$ is the current time step and the ego vehicle has observed the trajectory of the NV for previous $r$ time steps. $h$ captures the NV vehicle dynamics and $g_j$ are the convex inequality constraints in the NV's optimization problem. Other expressions have the same meaning as before and we denote trajectory data time step in the superscript.

With this, the residuals to be minimized are

\begin{subequations}\label{residuals}
    \begin{gather}
        r_{stat}(\alpha, \lambda, \nu) = \Sigma_{i=k-r}^{k}[\nabla J_{NV} + \Sigma_j \lambda_j\nabla g_j + \Sigma_j \nu_j\nabla h_j]
        \\ r_{comp}(\lambda) = \Sigma_{i=k-r}^{k}\lambda_j g_j, \;\;\; j=1,...,p
    \end{gather}
\end{subequations}
Note that in our formulation, we fit a problem (\ref{fit problem}) such that only the state trajectory data of NV is required. Specifically, problem (\ref{fit problem}) is carefully chosen such that equation (\ref{residuals}) results in the residuals containing only the state trajectory terms. This is vital for practical implementation since the control data of NV may not be easily available.

Finally, the following convex optimization problem is solved to impute the cost function based on the trajectory data.
\begin{subequations}
    \begin{gather}
        \min_{\alpha} \Vert r_{stat} \Vert _2 ^2 + \Vert r_{comp} \Vert _2 ^2
        \\ s.t. \;\;\; \alpha_s, \alpha_v, \alpha_a \geq 0
        \\ \lambda_j \geq 0
        \\ \alpha_s + \alpha_v + \alpha_a = 1
    \end{gather}
\end{subequations}

\section{SIMULATION RESULTS}\label{simulation}
We test our method in simulation. A 2-lane highway with on-ramp is considered where the ego vehicle (blue) needs to negotiate a merge with the NV (red). The aiMPC is invoked at a configuration where which vehicle should yield
is ambiguous. As discussed in literature, in the cases where the two cars are close in longitudinal direction and have similar speed, successfully
executing the merging maneuver is a very challenging task \cite{wang2020game}. This challenge stems from the unknown future trajectory of the NV and its unknown nature. We test various cases with NV's nature defined by the position, velocity and acceleration weights: $q_s, q_v$ and $q_a$, respectively, in its cost function. Qualitatively, for instance, when $q_v = 1$ the NV is \textit{aggressive} while when $q_a=1$ the NV is \textit{conservative} in its driving.

The MPC and imputation problems are solved using GUROBI\textsuperscript{\texttrademark} on a PC with Intel\textsuperscript{\textregistered} Core\textsuperscript{\texttrademark} i7-10750H CPU @ 2.60GHz and 8 GB RAM.

\begin{table}[h!]
\centering
\caption{Simulation parameters}
\begin{tabular}{c c} 
 \hline
 Simulation Parameter & Value \\ [0.5ex] 
 \hline\hline
 Simulation time & $8\;s$\\ 
 \hline
 Sampling time & $400\;ms$ \\
 \hline
 $N$ (Ego Horizon) & $15\;(6 s)$ \\
 \hline
 $T$ (NV Horizon) & $3\;(1.2 s)$ \\
 \hline
 $r$ (Trajectory Observation) & $3\;(1.2 s)$ \\ [1ex] 
 \hline
\end{tabular}
\label{sim param}
\end{table}

\begin{figure}[h!]
\begin{tikzpicture}
\filldraw [fill=white, draw=red, loosely dashed] (0,0) rectangle (3,-3);
\node at (1.5,-0.5) [red] {\small NV};
\node at (1.5,-1.5) [draw=black] (QCQP) {\footnotesize QCQP MPC};

\filldraw [fill=white, draw=blue, loosely dashed] (4.5,0) rectangle (8.5,-4);
\filldraw [fill=white, draw=black] (5,-1) rectangle (8,-3.25);
\node at (6.5,-0.5) [blue] {\small Ego};
\node at (6.5,-1.5) [draw=black] (BE) {\footnotesize Behavior Estimator};
\node at (6.5,-2.5) [draw=black] (JMPC) {\footnotesize Joint MPC};
\node at (6.5,-3) [blue] {\footnotesize aiMPC};

\draw [thick,->,>=stealth] (BE) -- node[anchor=west] {$\alpha$} (JMPC);
\draw [thick,->,>=stealth] (JMPC) -| node[anchor=east] {$X_{ego}$} (QCQP);
\draw [thick,->,>=stealth] (QCQP) -- node[anchor=south] {$X_{NV}$} (BE);
\end{tikzpicture}

\caption{The simulation process flow of information}
\end{figure}
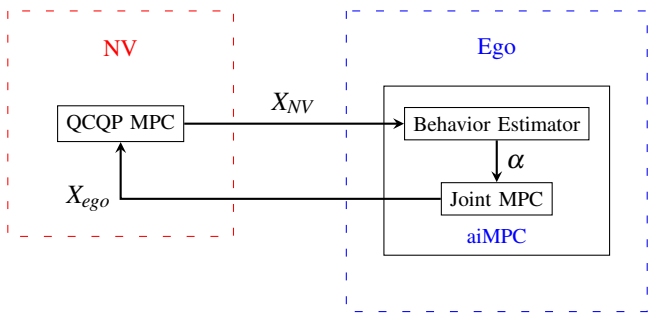

\subsection{Comparison with constant velocity and constant acceleration models}
We compare our method to constant velocity and constant acceleration models of NV. These are the baseline cases where the ego vehicle assumes that the NV remains at the current observed velocity and acceleration respectively, over the MPC horizon, at each time instant. The results of an edge case scenario (Table \ref{sim init}) are shown.

\begin{table}[h!]
\centering
\caption{Simulation A conditions}
\resizebox{0.5\textwidth}{!}{
\begin{tabular}{c c c c c c} 
 \hline
 NV Nature &  $(q_s, q_v, q_a)$ & $v_{0,ego}[m/s]$ & $v_{0,NV}[m/s]$ & Method & Result \\ [0.5ex] 
 \hline\hline
 Aggressive & (0, 1, 0) & 10 & 12 & \makecell{Constant velocity \\ non-interactive} & Ego unable to merge\\[0.5ex]
 \hline
 Aggressive & (0, 1, 0) & 10 & 12 & \makecell{Constant acceleration \\ non-interactive} & Ego unable to merge\\[0.5ex]
 \hline
 Aggressive & (0, 1, 0) & 10 & 12 & aiMPC & \makecell{Ego merges behind \\ $46.6 \%$ less hindraance}\\[0.5ex]
 \hline
\end{tabular}}
\label{sim init}
\end{table}

\begin{figure}[h!]
    \centering
    \includegraphics[width=0.5\textwidth]{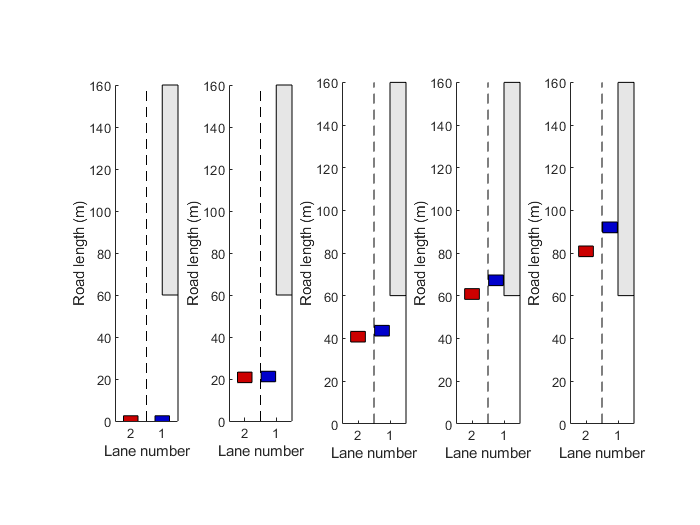}
    \caption{Constant velocity model for NV (Red): Ego (Blue) unable to merge}
    \label{const v NV}
\end{figure}

\begin{figure}[h!]
    \centering
    \includegraphics[width=0.5\textwidth]{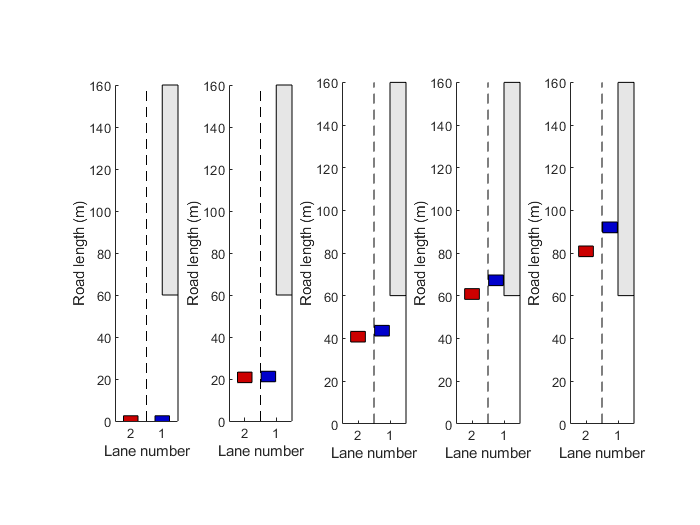}
    \caption{Constant acceleration model for NV (Red): Ego (Blue) unable to merge}
    \label{const a NV}
\end{figure}

\begin{figure}[h!]
    \centering
    \includegraphics[width=0.5\textwidth]{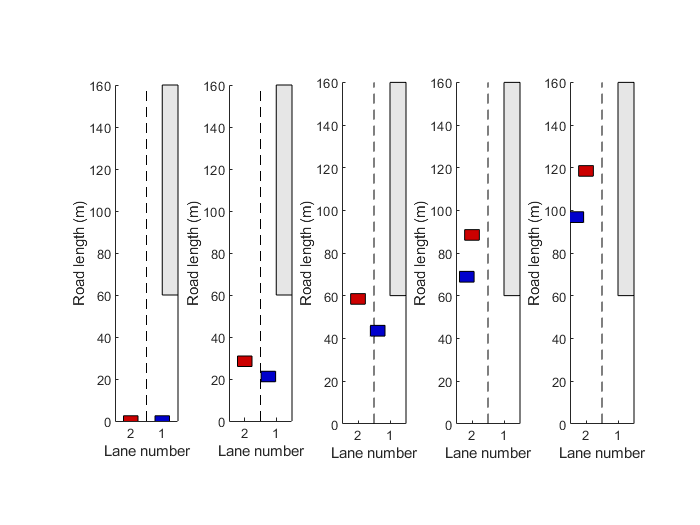}
    \caption{aiMPC: Ego (Blue) merges behind NV (Red) and reduces hindrance to NV}
    \label{merge A}
\end{figure}

\begin{figure}[h!]
    \centering
    \includegraphics[width=0.5\textwidth]{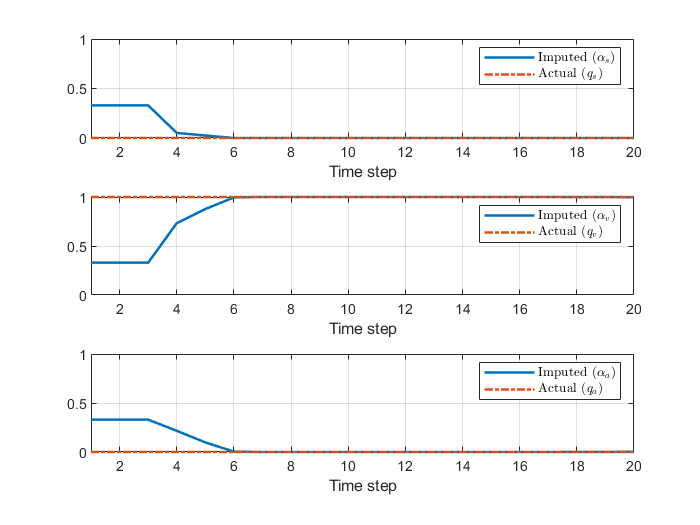}
    \caption{aiMPC imputation results: initialized with equi-weighted cost and adapted online based on trajectory observations}
    \label{imputation 1}
\end{figure}

We observe that the ego vehicle is unable to merge in the baseline cases and also causes hindrance for the NV (Fig. \ref{const v NV}, Fig. \ref{const a NV}). On the other hand, motion planning via aiMPC results in successful merging and reduced hindrance (Fig. \ref{merge A}). The hindrance to NV is reduced by $46.6 \%$ (travel distance increased), which reflects socially compliant driving.
This highlights the utility of \textit{interactive} motion planning. Fig. \ref{imputation 1} shows the online imputation results of NV's cost by the ego which are utilized in the interactive planning.

\subsection{Comparison with Non-adaptive MPC}
Next, we test the significance of estimating the nature of NV via the imputation method presented in the paper. We perform two simulations: one with equi-weighted cost of NV which remains constant throughout, and the other with (adaptive) online imputation of the cost.

\begin{table}[h!]
\centering
\caption{Simulation B conditions}
\resizebox{0.5\textwidth}{!}{
\begin{tabular}{c c c c c c} 
 \hline
 NV Nature &  $(q_s, q_v, q_a)$ & $v_{0,ego}[m/s]$ & $v_{0,NV}[m/s]$ & Method & Result \\ [0.5ex] 
 \hline\hline
 Conservative & (0, 0, 1) & 11 & 10 & Non-adaptive & \makecell{Ego merges ahead \\ but jerky} \\[0.5ex]
 \hline
 Conservative & (0, 0, 1) & 11 & 10 & aiMPC & \makecell{Ego merges ahead \\ smoother}\\[0.5ex]
 \hline
\end{tabular}}
\label{sim init B}
\end{table}


\begin{figure}[h!]
    \centering
    \includegraphics[width=0.4\textwidth]{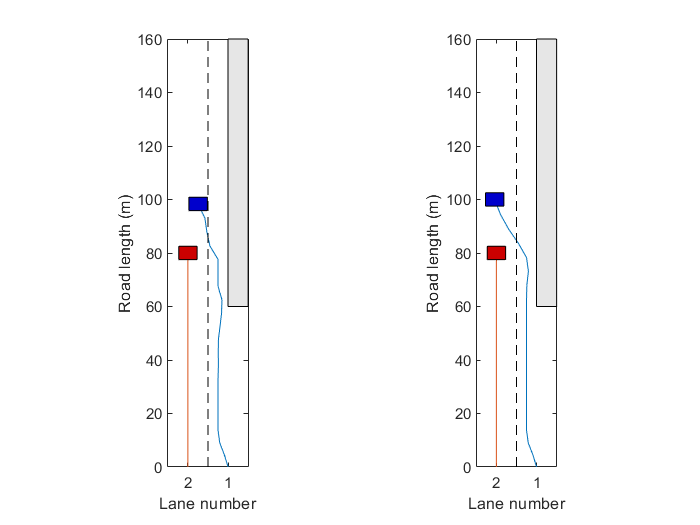}
    \caption{Non-adaptive cost (left) results in tentative motion while aiMPC (right) results in smoother motion}
    \label{fig: sim B}
\end{figure}


\begin{figure}[h!]
    \centering
    \includegraphics[width=0.4\textwidth]{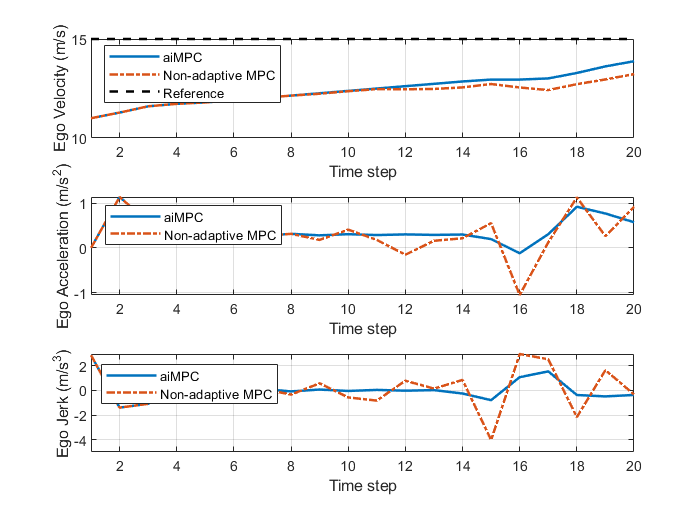}
    \caption{Ego is able to reach nearer to reference velocity and with lower jerk when the cost is adapted via online imputation}
    \label{motion compare}
\end{figure}

\begin{figure}[H]
    \centering
    \includegraphics[width=0.5\textwidth]{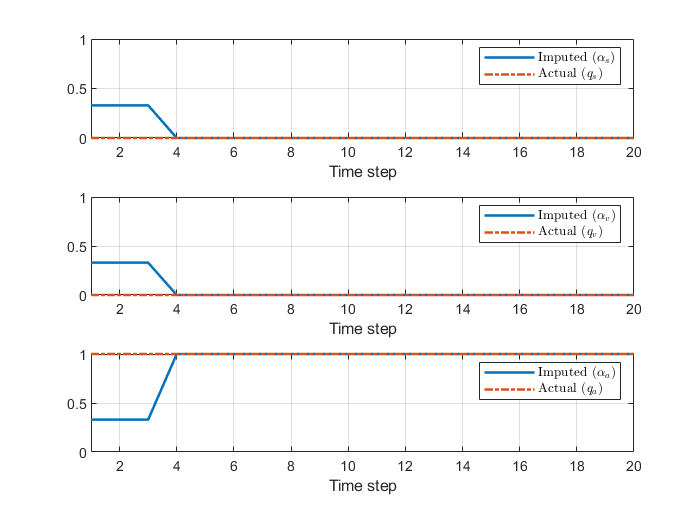}
    \caption{aiMPC imputation results: initialized with equi-weighted cost and adapted online based on trajectory observations}
    \label{imputation 2}
\end{figure}

In the first case (Fig. \ref{fig: sim B}), the ego merges ahead but the trajectory is not smooth. This tentative motion is due to the larger prediction error in the joint MPC because of the cost function mismatch. In the second case (Fig. \ref{fig: sim B}) the cost is imputed and adapted online. This leads to more accurate predictions in the joint MPC, resulting in more confident and smooth motion quantified through lower jerk (Fig. \ref{motion compare}). Fig. \ref{imputation 2} shows the online imputation results for this case.

\subsection{Moderate nature of NV}
We also test a mixed case where NV has a moderate nature and places equal weightage on \textit{aggressiveness} and \textit{conservativeness} i.e. $q_v = 0.5$ and $q_a = 0.5$.

\begin{table}[h!]
\centering
\caption{Simulation C conditions}
\resizebox{0.5\textwidth}{!}{
\begin{tabular}{c c c c c c} 
 \hline
 NV Nature &  $(q_s, q_v, q_a)$ & $v_{0,ego}[m/s]$ & $v_{0,NV}[m/s]$ & Method & Result \\ [0.5ex] 
 \hline\hline
 Moderate & (0, 0.5, 0.5) & 10 & 10 & Non-adaptive & Ego unable to merge \\[0.5ex]
 \hline
 Moderate & (0, 0.5, 0.5) & 10 & 10 & aiMPC & \makecell{Ego merges behind}\\[0.5ex]
 \hline
\end{tabular}}
\label{sim init C}
\end{table}


\begin{figure}[h!]
    \centering
    \includegraphics[width=0.4\textwidth]{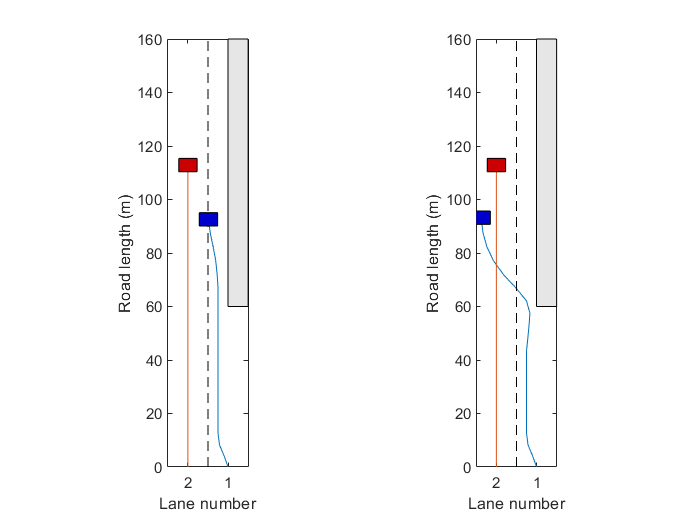}
    \caption{Nonadaptive MPC unable to merge in time (left) while aiMPC facilitates successful merging (right)}
    \label{fig: sim C}
\end{figure}

\begin{figure}[h!]
    \centering
    \includegraphics[width=0.5\textwidth]{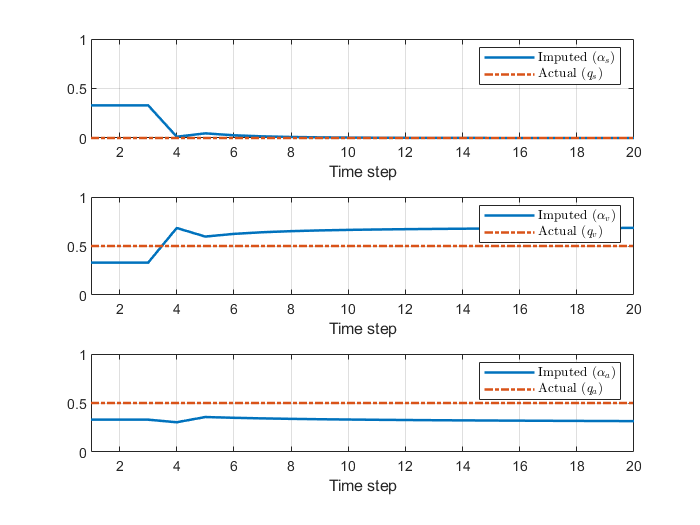}
    \caption{aiMPC does not converge to exact NV nature but the approximate imputation facilitates timely merging}
    \label{imputation C}
\end{figure}

In this case, the imputation cost does not converge exactly to the actual nature of NV (Fig. \ref{imputation C}), but the close approximation facilitates successful merging as shown in Fig. \ref{fig: sim C}.

\section{CONCLUSIONS}
We presented a new optimal control and inverse optimal control based framework for motion planning of automated vehicles interacting with neighboring vehicles. We focused on highway merging scenario due to the significant interactive challenges it presents. The proposed framework was tested in simulation and the results highlight the utility and significance of motion planning via interactive, online estimation based adaptive cost identifying MPC in highway merging.
Future work involves identification of higher fidelity cost functions and testing of the proposed framework in vehicle-in-the-loop tests.

\addtolength{\textheight}{-12cm}   








\bibliographystyle{ieeetr}
\bibliography{aiMPC}

\end{document}